\documentclass[twocolumn,
                showpacs,
                nofootinbib,
                nobibnotes,
                aps,
                superscriptaddress,
                amssymb,
                amsmath,
                floatfix,
                reprint,
                prl,
                notitlepage,
                showkeys,
                10pt,
		longbibliography,
		]{revtex4-1}

\usepackage{amssymb}
\usepackage{amsmath}
\usepackage{cprotect}
\usepackage{graphicx}
\usepackage{dcolumn}
\usepackage{url}
\usepackage[colorlinks=true,breaklinks=true,allcolors=blue]{hyperref}
\usepackage{tikz}
\usepackage{dsfont}
\usepackage{epsfig}
\usepackage{color}
\usepackage[ruled]{algorithm2e}
\usepackage{bbold}
\usepackage[binary-units=true]{siunitx}
\usepackage{glossaries}
\usepackage{tikz}
\usepackage{bm}
\usepackage{hhline}
\usepackage{spverbatim}
\usepackage{mhchem}
\usepackage{siunitx}

\makeatletter
\let\cat@comma@active\@empty
\makeatother

\newcommand{\Eq}[1]{Eq.\,\eqref{eq:#1}}
\newcommand{\Fig}[1]{Fig.~\ref{fig:#1}}

\begin{document}
\title{Variational Monte Carlo with Large Patched Transformers}

\author{Kyle Sprague}
\affiliation{Department of Physics, University of Ottawa, Ottawa, Ontario, K1N 6N5, Canada}
\author{Stefanie Czischek}
\email{stefanie.czischek@uottawa.ca}
\affiliation{Department of Physics, University of Ottawa, Ottawa, Ontario, K1N 6N5, Canada}

\date{\today}

\begin{abstract}
Large language models, like transformers, have recently demonstrated immense powers in text and image generation.
This success is driven by the ability to capture long-range correlations between elements in a sequence.
The same feature makes the transformer a powerful wavefunction ansatz that addresses the challenge of describing correlations in simulations of qubit systems.
Here we consider two-dimensional Rydberg atom arrays to demonstrate that transformers reach higher accuracies than conventional recurrent neural networks for variational ground state searches.
We further introduce large, patched transformer models, which consider a sequence of large atom patches, and show that this architecture significantly accelerates the simulations.
The proposed architectures reconstruct ground states with accuracies beyond state-of-the-art quantum Monte Carlo methods, allowing for the study of large Rydberg systems in different phases of matter and at phase transitions.
Our high-accuracy ground state representations at reasonable computational costs promise new insights into general large-scale quantum many-body systems.
\end{abstract}
\maketitle
%
%
%
\section{Introduction}
The advent of artificial neural network quantum states marks a turn in the field of numerical simulations for quantum many-body systems~\cite{Carleo2017, Torlai2018, Torlai2018a, Dawid2022, Carrasquilla2020, Carrasquilla2021}.
Since then, artificial neural networks are commonly used as a general wavefunction ansatz to find ground states of a given Hamiltonian~\cite{Hibat-Allah2020, Carleo2017, Czischek2022, Viteritti2022a}, to reconstruct quantum states from a set of projective measurements~\cite{Torlai2018, Torlai2018a, Neugebauer2020, Schmale2022, Torlai2019, Morawetz2021, Cha2022, Carrasquilla2019, Torlai2020}, or to model dynamics in open and closed quantum systems~\cite{Schmitt2020, Carleo2017, Nagy2019, Vicentini2019, Hartmann2019, Reh2021}.
The powers and limitations of different network architectures, such as restricted Boltzmann machines~\cite{Carleo2017, Torlai2018, Torlai2018a, Czischek2018, Torlai2019, Carrasquilla2019, Melko2019, Viteritti2022a}, recurrent neural networks (RNNs)~\cite{Hibat-Allah2020, Hibat-Allah2021, Morawetz2021, Czischek2022, Hibat-Allah2023}, or the PixelCNN~\cite{Sharir2020}, have been widely explored on several physical models.
In addition, modified network architectures~\cite{Valenti2022}, the explicit inclusion of symmetries~\cite{Hibat-Allah2022, Morawetz2021, Khandoker2023, Luo2023}, and the pre-training on a limited amount of measurement data~\cite{Bennewitz2022, Czischek2022} have shown improved performances.

A particularly promising choice are autoregressive neural networks such as the PixelCNN~\cite{Sharir2020} and RNNs~\cite{Hibat-Allah2020, Hibat-Allah2022, Morawetz2021, Czischek2022}, which can find ground states and reconstruct quantum states from data with high accuracies.
These models consider qubit systems in sequential order, providing an efficient wavefunction encoding.
However, these setups experience limitations for systems with strong correlations between qubits far apart in the sequence, which, for example, happens for two-dimensional qubit systems~\cite{Hibat-Allah2020, Czischek2022, Sharir2020}.

Similar to the RNN or PixelCNN approaches, transformer (TF) models~\cite{Vaswani2017} can be used as a wavefunction ansatz by considering a sequence of qubits~\cite{Zhang2023, Viteritti2022, Sharir2022, Cha2022, Ma2023, An2023, Glehn2022} or for simulating quantum dynamics~\cite{Carrasquilla2021a, Luo2022}.
Due to their non-recurrent nature and the ability to highlight the influence of specific previous sequence elements, TF models perform better at covering long-range correlations~\cite{Vaswani2017}, promising to overcome the limitations of RNNs and PixelCNNs~\cite{Viteritti2022, Sharir2022}.
In this work, we analyze the performance of the TF wavefunction ansatz for variational ground state searches and observe improved accuracies in the representation of quantum states compared to the RNN approach.

Inspired by the introduction of the vision transformer, which enables the efficient application of TF models for image processing and generation tasks~\cite{Dosovitskiy2021}, and by previous works in the field~\cite{Viteritti2022, Sharir2022, Hibat-Allah2023}, we study RNN and TF models that consider sequences of patches of qubits.
This approach reduces the sequence length and thus the computational cost significantly, while accurately capturing correlations within the patch.
For further improvements we introduce large, patched transformers (LPTF) consisting of a powerful patched TF model followed by a computationally efficient patched RNN that breaks large inputs into smaller sub-patches.
This architecture allows for an efficient consideration of large patches in the input sequence of the TF network, further reducing the sequence length.

We benchmark the LPTF architecture on two-dimensional arrays of Rydberg atoms, whose recently demonstrated experimental controllability makes them promising candidates for high-performance quantum computation and quantum simulation~\cite{Jaksch2000, Lukin2001, Endres2016, Barredo2018, Samajdar2020, Samajdar2021, Ebadi2021, Scholl2021, Xu2021, Miles2021, Kalinowski2022}.
Furthermore, quantum Monte Carlo methods can model Rydberg atom systems~\cite{Merali2021, Kalinowski2022}, and we use such simulations to determine the performance of different network models.

Analyzing different shapes and sizes of input patches, we demonstrate that LPTFs can represent ground states of Rydberg atom arrays with accuracies beyond the RNN ansatz and traditional quantum Monte Carlo simulations, while requiring reasonable computational costs.
Our results are consistent in different phases of matter and at quantum phase transitions.
While we show that LPTFs can significantly improve numerical investigations of the considered Rydberg models, the introduced network model can similarly be applied to general qubit systems.
The results presented in this work propose that the LPTF model can substantially advance numerical studies of quantum many-body physics.
%
%
\section{Results}
\subsection{Rydberg atom arrays}
\begin{figure*}
	\centering
	\includegraphics{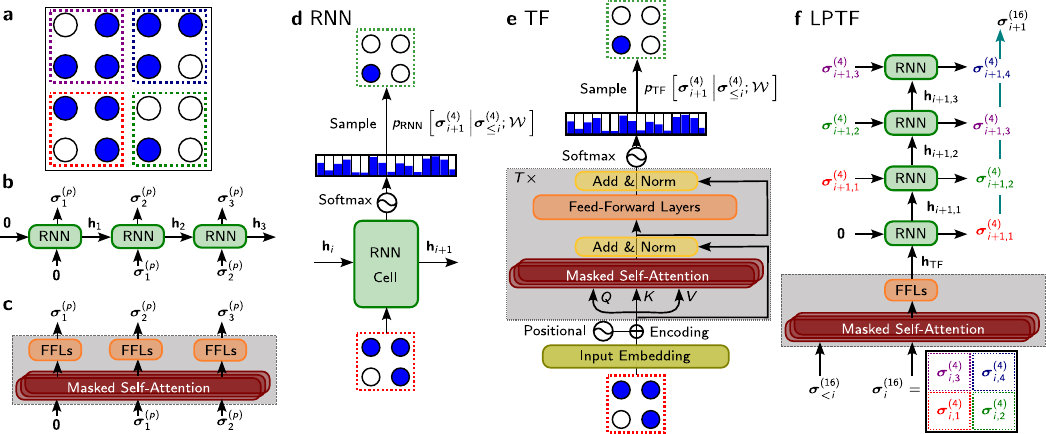}
	\caption{\textbf{Illustrating different network models.}
		\textbf{a}, Square lattice of $N=4\times 4$ Rydberg atoms randomly occupying the ground state (white) and the Rydberg state (blue). 
			Dash-colored squares indicate patches used as network inputs.
		\textbf{b}, Recurrent neural network (RNN) processing sequence.
			The RNN cell iteratively receives input sequence elements $\boldsymbol{\sigma}_i$ together with a hidden state.
			At each iteration, the output is used as the next input.
			The index $\left(p\right)$ denotes the input size in the patched RNN.
		\textbf{c}, Autoregressive transformer (TF) processing sequence, similar to the RNN in \textbf{b}.
			The multi-headed masked self-attention layer generates weighted connections to previous input sequence elements.
			For simplicity, we only include the feed-forward layers (FFLs) in the scheme.
		\textbf{d}, Single patched RNN iteration on inputs of patch size $p=2\times 2$ [indicated with $\left(4\right)$].
			A softmax function creates a probability distribution $p_{\mathrm{RNN}}$ over all possible patch states conditioned on previous sequence elements.
			The state of the next patch is sampled and used as next input state.
		\textbf{e}, Single patched TF iteration.
			The input patch is embedded into a state of dimension $d_{\mathrm{H}}$, and a positional encoding keeps track of the sequence order.
			The signal is sent into the transformer cell (gray) which we generate similar to~\cite{Vaswani2017} and apply $T$ times independently.
			The output of the transformer cells is used like in the patched RNN to sample the next input.
		\textbf{f}, Single large, patched transformer (LPTF) iteration for patch size $p=4\times 4$ [indicated with $\left(16\right)$] and sub-patch size $p_{\mathrm{s}}=2\times 2$.
			A patched TF model receives a large input patch, and the transformer cell output is propagated as a hidden state $\boldsymbol{h}_{\mathrm{TF}}$ to a patched RNN. 
			The patched RNN autoregressively constructs the input patch of the next LPTF iteration, reducing the output dimension.
		See Methods for more details on the network models.
	}
	\label{fig:1}
\end{figure*}
Rydberg atoms, which we use as a qubit model to benchmark our numerical approaches, can be prepared in the ground state $\left|\mathrm{g}\right\rangle$ and in a highly excited (Rydberg) state $\left|\mathrm{r}\right\rangle$~\cite{Jaksch2000, Lukin2001, Endres2016, Samajdar2020, Ebadi2021, Barredo2018, Scholl2021}.
We specifically consider the atoms arranged on square lattices of different system sizes, as illustrated in \Fig{1}\textbf{a}.

The system of $N=L\times L$ atoms is described by the Rydberg Hamiltonian~\cite{Jaksch2000, Lukin2001},
\begin{align}
	\hat{\mathcal{H}} &= -\frac{\Omega}{2}\sum_{i=1}^N\hat{\sigma}_i^x-\delta\sum_{i=1}^N\hat{n}_i+\sum_{i,j}V_{i,j}\hat{n}_i\hat{n}_j,
	\label{eq:Hamiltonian}
\end{align}
with the detuning $\delta$ and the Rabi oscillation with frequency $\Omega$ generated by an external laser driving.
Here we use the off-diagonal operator $\hat{\sigma}_i^x=\left|\mathrm{g}\right\rangle_i\left\langle\mathrm{r}\right|_i+\left|\mathrm{r}\right\rangle_i\left\langle\mathrm{g}\right|_i$ and the occupation number operator $\hat{n}_i=\left|\mathrm{r}\right\rangle_i\left\langle\mathrm{r}\right|_i$.
The last term in the Hamiltonian describes a van-der-Waals interaction between atoms at positions $\boldsymbol{r}_i$ and $\boldsymbol{r}_j$, with $V_{i,j}=\Omega R_{\mathrm{b}}^6 / \left|\boldsymbol{r}_i-\boldsymbol{r}_j\right|^6$, and Rydberg blockade radius $R_{\mathrm{b}}$. 
We further choose the lattice spacing $a=1$.
By tuning the free parameters in the Rydberg Hamiltonian, the system can be prepared in various phases of matter, separated by different kinds of phase transitions~\cite{Ebadi2021, Samajdar2020, Kalinowski2022, Samajdar2021, Miles2021}.
The Rydberg Hamiltonian is stoquastic~\cite{Bravyi2007}, resulting in a positive and real-valued ground-state wavefunction~\cite{Endres2016, Ebadi2021, Xu2021}.
More details on Rydberg atom arrays are provided in the Methods section.
\subsection{Recurrent neural networks and transformers}
Recurrent neural networks (RNNs) provide a powerful wavefunction ansatz that can variationally find ground state representations of quantum many-body systems~\cite{Hibat-Allah2020, Morawetz2021, Czischek2022, Hibat-Allah2022, Khandoker2023, Carrasquilla2021, Dawid2022, Hibat-Allah2023}.
For this, the possibility to naturally encode probability distributions in RNNs allows the representation of squared wavefunction amplitudes $|\Psi\left(\boldsymbol{\sigma}\right)|^2$.
Samples drawn from the encoded distribution correspond to state configurations that can be interpreted as outputs of projective measurements, as illustrated in \Fig{1}\textbf{d}.

To represent the wavefunction amplitudes $\Psi\left(\boldsymbol{\sigma}\right)=\langle\boldsymbol{\sigma}|\Psi\rangle$ of a qubit system, such as an array of Rydberg atoms, a sequential order is defined over the system.
Each atom is iteratively used as an input to the RNN cell, the core element of the network structure which we choose to be a Gated Recurrent Unit (GRU)~\cite{Cho2014} inspired by~\cite{Hibat-Allah2020, Carrasquilla2021}.
In addition, the RNN cell receives the state of internal hidden units as input. 
This state is adapted in each iteration and propagated over the input sequence, generating a memory effect.
The network output $p_{\mathrm{RNN}}\left(\sigma_i|\sigma_{<i};\mathcal{W}\right)$ at each iteration can be interpreted as the probability of the next atom $\sigma_i$ being in either the ground or the Rydberg state, conditioned on the configuration of all previous atoms $\sigma_{<i}$ in the sequence, with variational weights $\mathcal{W}$ in the RNN cell.
See \Fig{1}\textbf{d} and the Methods section for more details.
From this output, the state $\sigma_i$ of the next atom is sampled and used autoregressively as input in the next RNN iteration, as illustrated in \Fig{1}\textbf{b}~\cite{Hibat-Allah2020, Carrasquilla2021, Dawid2022}.
We then train the RNN such that it approximates a target state $\Psi\left(\boldsymbol{\sigma}\right)$,
\begin{align}
	\nonumber
	\Psi_{\mathrm{RNN}}\left(\boldsymbol{\sigma};\mathcal{W}\right)&=\sqrt{\prod_{i=1}^N p_{\mathrm{RNN}}\left(\sigma_i|\sigma_{<i};\mathcal{W}\right)}\\
	\nonumber
	& = \sqrt{p_{\mathrm{RNN}}\left(\boldsymbol{\sigma};\mathcal{W}\right)}\\
	&\approx \Psi\left(\boldsymbol{\sigma}\right).
\end{align}
While we focus on positive, real-valued wave functions in this work, RNNs can represent general wave functions by including complex phases as a second network output~\cite{Hibat-Allah2020}.
The global phase of the encoded state is then expressed as the sum over single-qubit phases.

The RNN has shown high accuracies for representing ground states of various quantum systems.
However, its sequential nature and the encoding of all information in the hidden unit state pose a challenge for capturing long-range correlations~\cite{Hibat-Allah2020, Morawetz2021, Czischek2022, Hibat-Allah2022, Khandoker2023, Carrasquilla2021, Dawid2022}.
Here we refer to correlations between atoms that appear far from each other in the RNN sequence but not necessarily in the qubit system.
Alternative autoregressive network models, such as the PixelCNN, experience similar limitations.
These models cover correlations via convolutions with a kernel of a specific size.
However, due to increasing computational costs, kernel sizes are commonly chosen rather small, so that the PixelCNN is as well limited to capturing only local correlations in qubit systems~\cite{Sharir2020}.
Specific RNN structures that better match the lattice structure in the considered model, such as two-dimensional RNNs for two-dimensional quantum systems, can overcome this limitation~\cite{Hibat-Allah2020, Hibat-Allah2022, Khandoker2023, Hibat-Allah2023}.
An alternative approach to improve the representation of long-range correlations is to use transformer (TF) architectures as a wavefunction ansatz~\cite{Cha2022, Sharir2022, Viteritti2022, Zhang2023}.
These provide a similar autoregressive behavior but do not have a recurrent setup and naturally capture all-to-all interactions~\cite{Vaswani2017}.

While autoregressively using the states of individual atoms as sequential input similar to the RNN, a masked self-attention layer in the TF setup provides trained connections to all previous elements in the sequence~\cite{Vaswani2017}.
See \Fig{1}\textbf{e} and the Methods section for more details.
These trainable connections generate all-to-all interactions between the atoms in the system and allow the highlighting of high-impact connections or strong correlations.
This setup thus proposes to represent strongly correlated quantum systems with higher accuracy than the RNN model~\cite{Cha2022, Sharir2022, Viteritti2022}.
As illustrated in \Fig{1}\textbf{c}, the TF model outputs probability distributions which provide an autoregressive wavefunction ansatz  $\Psi_{\mathrm{TF}}\left(\boldsymbol{\sigma};\mathcal{W}\right)=\sqrt{p_{\mathrm{TF}}\left(\boldsymbol{\sigma};\mathcal{W}\right)}$~\cite{Cha2022, Sharir2022, Viteritti2022}, as further explained in the Methods section.
Similarly to the RNN, the TF network can represent complex-valued wave functions by adding a second output representing the single-qubit phases~\cite{Hibat-Allah2020}.
\begin{figure}
	\centering
	\includegraphics{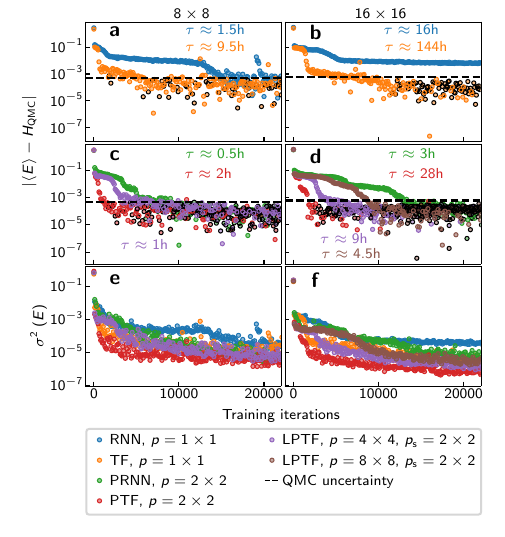}
	\caption{\textbf{Performance of different network architectures on Rydberg atom arrays.}
	\textbf{a, b} Absolute energy difference between $\langle E\rangle$ [\Eq{E_exp}] from $N_{\mathrm{s}}=512$ recurrent neural network (RNN, blue) and transformer (TF, orange) samples and $H_{\mathrm{QMC}}$ from $N_{\mathrm{s}}=7\times 10^4$ quantum Monte Carlo (QMC) samples as a function of network training iterations.
	The black dashed line denotes the QMC uncertainty and black-edged data points show the absolute value of energies below QMC results.
	$\tau$ is the total runtime in hours ($\mathrm{h}$) for $2\times 10^4$ training iterations using the network in the corresponding color (blue for RNN, orange for TF) on NVIDIA Tesla P100 GPUs.
	For the QMC, the total runtimes were obtained as $\tau=18\mathrm{h}$ for $N=8\times 8$ atoms and $\tau=24\mathrm{h}$ for $N=16\times 16$ atoms for $N_{\mathrm{s}}=10^4$ samples on a single CPU.
	\textbf{c, d} Same as \textbf{a, b} for the patched RNN (PRNN, green) and the patched TF model (PTF, red) with patch size $p=2\times 2$, and for the large, patched transformer (LPTF) approach (purple) with patch size $p=4\times 4$ and sub-patch size $p_{\mathrm{s}}=2\times 2$.
	For $N=16\times 16$ atoms, we further show LPTF results with patch size $p=8\times 8$ (brown).
	\textbf{e, f} Variances $\sigma^2\left(E\right)$ of energy expectation values for the network architectures considered in \textbf{a}-\textbf{d}.
	}
	\label{fig:2}
\end{figure}

In \Fig{2}\textbf{a},\textbf{b}, we compare the performance of RNNs (blue) and TFs (orange) when representing ground states of Rydberg arrays with $N=8\times 8$ (\textbf{a}) and $N=16\times 16$ (\textbf{b}) atoms.
Here we fix $R_{\mathrm{b}}=7^{1/6}\approx 1.383$ and $\Omega=\delta=1$, which brings us into the vicinity of the transition between the disordered and the striated phase~\cite{Ebadi2021}.
We variationally train the network models by minimizing the energy expectation value, corresponding to a variational Monte Carlo method~\cite{Melko2019, Carrasquilla2021, Becca2017, Dawid2022}, see Methods section.
If not stated otherwise, the energy expectation values in this work are evaluated on $N_{\mathrm{s}}=512$ samples generated from the network, which we consider in mini batches of $K=256$ samples.
We obtained satisfactory results with $d_{\mathrm{H}}=128$ hidden neurons in the RNN and the equivalent embedding dimension $d_{\mathrm{H}}=128$ in the TF model.
To benchmark the performance of the two models, we show the difference between the ground state energies $H_{\mathrm{QMC}}$ obtained from quantum Monte Carlo (QMC) simulations at zero temperature~\cite{Merali2021}, and the energy expectation value,
\begin{align}
	\langle E\rangle &= \frac{1}{N_{\mathrm{s}}}\sum_{s=1}^{N_{\mathrm{s}}}H_{\mathrm{loc}}\left(\boldsymbol{\sigma}_{s}\right),
	\label{eq:E_exp}
\end{align}
extracted from network samples $\boldsymbol{\sigma}_s$.
Here we use the local energy,
\begin{align}
	H_{\mathrm{loc}}\left(\boldsymbol{\sigma}_s\right) &= \frac{\langle \boldsymbol{\sigma}_s|\hat{\mathcal{H}}|\Psi_{\mathcal{W}}\rangle}{\langle\boldsymbol{\sigma}_s|\Psi_{\mathcal{W}}\rangle},
	\label{eq:E_loc}
\end{align}
with $|\Psi_{\mathcal{W}}\rangle$ denoting the wavefunction encoded in either the RNN or the TF network, as discussed in the Methods section.
In the QMC simulations, we use the stochastic series expansion approach presented in~\cite{Merali2021} and evaluate the expectation value on $N_{\mathrm{s}}=7\times 10^4$ samples generated from seven independent sample chains.
Both system sizes show that TFs converge to the ground state energy within fewer training iterations than the RNN.
Additionally, for the larger system in \Fig{2}\textbf{b}, TFs outperform RNNs significantly and reach higher accuracies in the ground state energy.
This result demonstrates the expected improved performance.

We, however, also find that this enhancement comes at the cost of increased computational runtimes $\tau$ in hours ($\mathrm{h}$) for $2\times 10^4$ training iterations.
With $\tau\approx 1.5\mathrm{h}$ and $\tau\approx 16\mathrm{h}$ for $N=8\times 8$ and $N=16\times 16$ atoms, RNNs process much faster than TFs with $\tau\approx9.5\mathrm{h}$ and $\tau\approx144\mathrm{h}$, respectively.
\Fig{2}\textbf{a},\textbf{b} suggest stopping the TF training after fewer iterations due to the faster convergence, but the computational runtime is still too long to allow scaling to large system sizes.

We obtained QMC runtimes as $\tau\approx 18\mathrm{h}$ for $N=8\times 8$ and $\tau\approx 24\mathrm{h}$ for $N=16\times 16$ for a single run generating $N_{\mathrm{s}}=10^4$ samples, showing a more efficient scaling with system size than the network simulations.
This behavior can be understood when considering the scaling of the computational cost for generating an individual sample, which is $\mathcal{O}\left(N\right)$ for the RNN and QMC, and $\mathcal{O}\left(N^2\right)$ for the TF.
In addition, the network models need to evaluate energy expectation values in each training iteration, which comes at complexity $\mathcal{O}\left(N^2\right)$ for the RNN and at complexity $\mathcal{O}\left(N^3\right)$ for the TF, see Methods for more details.
However, due to its non-recurrent setup, the TF enables a parallelization of the energy expectation value evaluation, which is not possible for the RNN ansatz, as further discussed in the Methods.
The computational complexity for QMC scales as $\mathcal{O}\left(N\right)$ for both sampling and energy evaluation~\cite{Merali2021}.
Thus, while the QMC requires longer runtimes than the RNN for small system sizes, it is expected to outperform both the RNN and the TF for larger systems.
\subsection{Patched inputs}
To address the exceeding computational runtime of TF models, we take inspiration from the vision transformer~\cite{Dosovitskiy2021} and consider patches of atoms as inputs to both considered network architectures, as illustrated in \Fig{1}\textbf{d},\textbf{e}.
This reduces the sequence length to $N/p$ elements for patch size $p$, leading to a sampling complexity of $\mathcal{O}\left(N/p\right)$ for the patched RNN and $\mathcal{O}\left(N^2/p^2\right)$ for the patched TF model, as well as an energy evaluation complexity of $\mathcal{O}\left(N^2/p\right)$ and $\mathcal{O}\left(N^3/p^2\right)$, respectively.

We first use patches of $p=2\times 2$ atoms.
The network output is then a probability distribution over the $2^p=16$ states the atoms in the patch can take, from which the next patch is sampled and used autoregressively as input in the following iteration.
As demonstrated in previous works~\cite{Sharir2022, Viteritti2022, Hibat-Allah2023}, this significantly reduces the computational runtime due to the shorter sequence length.
In addition, we expect it to capture correlations between neighboring atoms with higher accuracies by directly encoding them in the output probabilities.
The patched models can also be modified to include complex phases as a second network output, which then correspond to the sum of phases of individual qubits in the patch~\cite{Hibat-Allah2020}.

\Fig{2}\textbf{c} and \textbf{d} show the results for the same $N=8\times 8$ and $N=16\times 16$ atom Rydberg array ground states as in panels $\textbf{a}$ and $\textbf{b}$, using the patched RNN (green) and the patched TF setup (red) with $p=2\times 2$.
The network hyperparameters are the same as in the RNN and the TF network in \textbf{a} and \textbf{b}.
The computational runtime reduces significantly to $\tau\approx0.5\mathrm{h}$ and $\tau\approx3\mathrm{h}$, using the patched RNN and the patched TF model for $N=8\times 8$ atoms, and to $\tau\approx2\mathrm{h}$ and $\tau\approx28\mathrm{h}$, respectively, for $N=16\times 16$ atoms.
Convergence further happens within fewer training iterations than for the unpatched networks, and all representations reach energy values within the QMC errors.
We even observe energies below the QMC results, which always remain within the QMC uncertainties and thus do not violate the variational principle which we expect to be satisfied for the number of samples we use to evaluate energy expectation values and for the small variances we observe~\cite{Becca2017}.
These energies propose that the patched networks find the ground state with better accuracy than the QMC simulations using $N_{\mathrm{s}}=7\times 10^4$ samples.
The QMC accuracy can be further increased by using more samples, where the uncertainty decreases as $\propto 1/\sqrt{N_{\mathrm{s}}}$ for uncorrelated samples~\cite{Merali2021}.
However, samples in a single QMC chain are correlated, resulting in an uncertainty scaling $\propto \sqrt{\tau_{\mathrm{auto}}/N_{\mathrm{s}}}$ with autocorrelation time $\tau_{\mathrm{auto}}$ depending on the evaluated observable~\cite{Merali2021}.
Even though the computational cost of QMC scales linearly with the sample chain size $N_{\mathrm{s}}$ and is thus more efficient than the RNN or the TF approach, which require the generation of $N_{\mathrm{s}}$ samples in each training iteration, we found that reaching higher QMC precisions comes at runtimes that exceed the patched RNN and the patched TF due to long autocorrelation times for large system sizes.
\subsection{Large, patched transformers}
Based on the results with $p=2\times 2$, we expect even shorter computational runtimes and higher representation accuracies from larger patch sizes.
However, as illustrated in \Fig{1}\textbf{d},\textbf{e}, the network output dimension scales exponentially with the input patch size, encoding the probability distribution over all possible patch states.
This output scaling leads to the sampling cost scaling as $\mathcal{O}\left(2^pN/p\right)$ for the patched RNN and as $\mathcal{O}\left(N^2/p^2 +2^pN/p\right)$ for the patched TF network, as well as energy evaluation costs scaling as $\mathcal{O}\left(2^pN^2/p\right)$ and $\mathcal{O}\left(N^3/p^2+2^pN^2/p\right)$, respectively, see Methods.
A hierarchical softmax approach is often used in image processing to efficiently address this exponential scaling~\cite{Morin2005}.
Here we introduce large, patched transformers (LPTFs) as an alternative way to enable efficient patch size scaling.

As shown in \Fig{1}\textbf{f}, the LPTF model uses a patched TF setup and passes the TF state into a patched RNN as the initial hidden state.
The patched RNN splits the input patch into smaller sub-patches of size $p_{\mathrm{s}}=2\times 2$, reducing the output of the LPTF model to the probability distribution over the $2^{p_{\mathrm{s}}}=16$ sub-patch states, as further discussed in the Methods.
The sampling complexity for this model is reduced to $\mathcal{O}\left(N^2/p^2+2^{p_{\mathrm{s}}}N/p_{\mathrm{s}}\right)$ and the energy evaluation complexity takes the form $\mathcal{O}\left(N^3/p^2+2^{p_{\mathrm{s}}}N^2/p_{\mathrm{s}}\right)$, as derived in the Methods section.
Generally, we can use both the patched RNN and the patched TF architecture as base network and subnetwork.
We choose this setup here to combine the high accuracies reached with the patched TF network for large system sizes with the computational efficiency of the patched RNN, which can still accurately represent small systems (see \Fig{2}\textbf{a}).
Being a combination of a TF network and an RNN, the LPTF can similarly be modified to include complex phases as a second network output.

In \Fig{2}\textbf{c},\textbf{d}, we compare the performance of the LPTF model to the previously considered network architectures, where we choose $p=4\times 4$ (purple) and $p=8\times 8$ (brown), with $p_{\mathrm{s}}=2\times 2$, using the same hyperparameters for all networks.
These models require more training iterations than the patched TF architecture to converge but reach accuracies comparable to the patched RNN and the patched TF network.
Even though more training iterations are required, the computational runtimes are reduced to $\tau\approx1\mathrm{h}$ for $N=8\times 8$, $p=4\times 4$, as well as $\tau\approx9\mathrm{h}$ and $\tau\approx4.5\mathrm{h}$ for $N=16\times 16$ with $p=4\times 4$ and $p=8\times 8$, respectively.
Thus, overall, we obtain convergence within shorter computational runtime.

The observed runtimes are also shorter than QMC runs, even though QMC is expected to outperform the network models for large system sizes due to the linear scaling of computational costs with $N$.
However, QMC is based on the generation of a chain of correlated samples.
For large system sizes, autocorrelation times between samples in the chain increase and the ergodicity of the sampling process is not necessarily guaranteed~\cite{Merali2021}.
Since these limitations do not arise for the exact sampling process in autoregressive ANN methods~\cite{Hibat-Allah2020, Sharir2020}, computationally efficient architectures such as the LPTF are still promising candidates for accurate studies of large quantum many-body systems.

\Fig{2}\textbf{e} and \textbf{f} show the variances $\sigma^2\left(E\right)$ of the energy expectation values obtained with all considered network architectures.
As expected~\cite{Hibat-Allah2020}, they decrease to zero when converging to the ground state energies.
This behavior confirms the accurate ground state reconstruction, while the smoothness of all curves demonstrates stable training processes.
\begin{figure}
	\centering
	\includegraphics{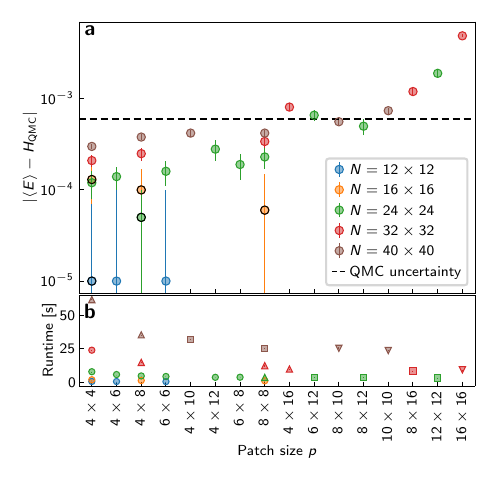}
	\caption{\textbf{Patch size scaling in large, patched transformers (LPTFs).}
	\textbf{a}, Absolute energy difference between $\langle E\rangle$ [\Eq{E_exp}] evaluated with $N_{\mathrm{s}}=512$ large, patched transformer (LPTF) samples and $H_{\mathrm{QMC}}$ evaluated with $N_{\mathrm{s}}=7\times 10^4$ quantum Monte Carlo (QMC) samples for different system sizes $N$ (colors) as a function of the patch size $p$.
	Error bars denote the standard error of the sampling mean.
	The black dashed line indicates the QMC uncertainty, while black-edged data points show absolute values of energies below the QMC results.
	\textbf{b}, Computational runtime per training iteration on NVIDIA Tesla P100 GPUs averaged over $2\times 10^4$ iterations of LPTFs as in \textbf{a}.
	Different shapes denote different mini-batch sizes $K$, with $K=256$ for circles, $K=128$ for up-pointing triangles, $K=64$ for squares, and $K=32$ for down-pointing triangles, see Methods.
	Error bars are smaller than the data points.	
	}
	\label{fig:3}
\end{figure}

We can further increase the patch size $p$ in the LPTF architecture, from which we expect even shorter runtimes.
However, this also increases the patch size that needs to be reconstructed with the patched RNN.
We thus expect the accuracy to decrease for large $p$ if we keep $p_{\mathrm{s}}=2\times 2$ fixed.
\Fig{3}\textbf{a} shows the energy difference between QMC and LPTF simulations for ground states of Rydberg arrays with $N=12\times 12$ up to $N=40\times 40$ atoms.
We keep the parameters at $R_{\mathrm{b}}=7^{1/6}$, $\delta=\Omega=1$, and evaluate the QMC energies on $N_{\mathrm{s}}=7\times 10^4$ samples from seven independent chains~\cite{Merali2021}, where the computational cost for QMC scales as $\mathcal{O}\left(N\right)$ with the system size.
Each LPTF data point corresponds to an average over training iterations $19{,}000$ to $20{,}000$ of ten independently trained networks with the same setup as for \Fig{2}.
We vary the input patch size between $p=4\times 4$ and $p=16\times 16$, where we also consider rectangular-shaped patches while fixing $p_{\mathrm{s}}=2\times 2$.
We ensure that the system size always divides by the input patch size.

As expected, the energy accuracies decrease with increasing patch size, which might result from the limited representational power of the patched RNN for large input $p$ and small $p_{\mathrm{s}}$ and from the increased amount of information that is encoded in each network iteration.
We find accuracies below the QMC uncertainty for up to $p=8\times 8$, which still proposes a significant speed-up compared to single-atom inputs in the TF model, see \Fig{2}\textbf{d}.
\Fig{3}\textbf{b} shows the computational runtimes of single training iteration steps for the different patch and system sizes.
Each data point shows an average over $2\times 10^4$ training iterations in a single network.
We find a rapid decrease in computation times for small patches while we observe convergence to steady times for larger patches.
This behavior results from the increased memory required by larger patch sizes, which forces us to decrease the mini-batch size $K$ of samples for the energy evaluation, see Methods.
Smaller mini-batch sizes lead to increased runtimes, which compete with the acceleration from the reduced sequence lengths.

We cannot find a conclusive dependence on the patch shape, with rectangular patches showing a similar behavior as squared patches.
Thus, the only important factor is the overall patch size, and we conclude that input patches around $p=8\times 8$ atoms provide a good compromise with reduced computation times and high energy accuracies.
\subsection{Phases of matter in Rydberg atom arrays}
\begin{figure}
	\centering
	\includegraphics{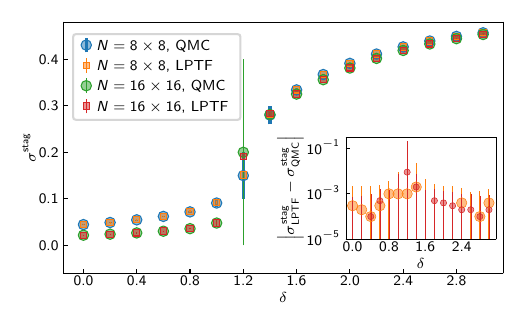}
	\caption{\textbf{Staggered magnetization as order parameter.}
	Staggered magnetization $\sigma^{\mathrm{stag}}$ [\Eq{2}] obtained with large, patched transformers (LPTFs, squares) and with quantum Monte Carlo (QMC, circles) for Rydberg arrays with $N=8\times 8$ (blue, orange) and $N=16\times 16$ (green, red) atoms when driving the detuning $\delta$ across the transition between the disordered ($\delta\lessapprox 1.2$) and the checkerboard ($\delta\gtrapprox 1.2$) phase.
	For $N=8\times 8$, we use patch size $p=4\times 4$, while $p=8\times 8$ for $N=16\times 16$.
	LPTF data is averaged over training iterations $11{,}000$ to $12{,}000$ of five independent networks with $N_{\mathrm{s}}=512$ samples.
	QMC data is evaluated on $N_{\mathrm{s}}=7\times 10^5$ samples.
	Error bars denote the standard error of the sampling mean, where autocorrelation times are considered in the QMC samples~\cite{Merali2021}.
	\textit{Inset:} Absolute difference between the LPTF and QMC data in the main plot for $N=8\times 8$ (orange) and $N=16\times 16$ (red).
	Error bars denote the standard error of the sampling mean.
	}
	\label{fig:4}
\end{figure}
We now explore the performance of LPTFs at different points in the Rydberg phase diagram by varying the detuning from $\delta=0$ to $\delta=3$ and fixing $R_{\mathrm{b}}=3^{1/6}\approx 1.2$, $\Omega=1$. 
With this, we drive the system over the transition between the disordered and the checkerboard phase~\cite{Ebadi2021, Samajdar2020}.
The order parameter for the checkerboard phase is given by the staggered magnetization~\cite{Samajdar2020},
\begin{align}
	\sigma^{\mathrm{stag}} &= \left\langle \left|\sum_{i=1}^N\left(-1\right)^i\frac{n_i-1/2}{N}\right|\right\rangle,
	\label{eq:2}
\end{align}
where $i$ runs over all $N=L\times L$ atoms and $n_i=|r\rangle_i\langle r|_i$ is the occupation number operator acting on atom $i$.
The expectation value denotes the average over sample configurations generated via QMC or from the trained network.

\Fig{4} shows the staggered magnetization when tuning $\delta$ over the phase transition, where we compare LPTF and QMC simulations.
The QMC data points show the average of $N_{\mathrm{s}}=7\times 10^5$ samples generated from seven independent chains~\cite{Merali2021}.
The LPTF data is averaged over training iterations $11{,}000$ to $12{,}000$ of five independently trained networks.
We look at systems with $N=8\times 8$ and $N=16\times 16$ atoms, choosing patch sizes $p=4\times 4$ and $p=8\times 8$, respectively, with $p_{\mathrm{s}}=2\times 2$.

The LPTF model captures the phase transition accurately for both system sizes, overlapping closely with the QMC results for all $\delta$ and showing small uncertainties.
In the inset in \Fig{4}, we plot the absolute difference between the staggered magnetizations obtained with QMC and LPTFs for both system sizes.
The most challenging regime to simulate is at $\delta\approx 1.2$, where we find the phase transition in the main panel.
Here the observed difference is $\sim 10^{-2}$, demonstrating the high accuracies reachable with the LPTF approach.
In the vicinity of the phase transition, the QMC uncertainties increase.
This behavior is related to long autocorrelation times $\tau_{\mathrm{auto}}$ in the individual sample chains and the uncertainty scaling as $\propto \sqrt{\tau_{\mathrm{auto}}/N_{\mathrm{s}}}$~\cite{Merali2021}.
The errors in the LPTF simulations remain small here, demonstrating a consistent and accurate outcome in all independent networks.
%
%
\section{Discussion}
We explored the power of transformer (TF) models~\cite{Vaswani2017} in representing ground states of two-dimensional Rydberg atom arrays of different sizes by benchmarking them on quantum Monte Carlo simulations~\cite{Merali2021}.
Our work provides a careful performance comparison of TF models with a recurrent neural network (RNN) wavefunction ansatz~\cite{Hibat-Allah2020, Carrasquilla2021, Dawid2022}, showing that TFs reach higher accuracies, especially for larger system sizes, but require longer computational runtimes.
We accelerate the network evaluation using patches of atoms as network inputs inspired by the vision transformer~\cite{Dosovitskiy2021} and demonstrate that these models significantly improve computational runtime and reachable accuracies.

Based on the obtained results, we introduce large, patched transformers (LPTFs), which consist of a patched TF network whose output is used as the initial hidden unit state of a patched RNN.
This model enables larger input patch sizes which are broken down into smaller patches in the patched RNN, keeping the network output dimension at reasonable size.

The LPTF models reach accuracies below the QMC uncertainties for ground states obtained with a fixed number of samples, while requiring significantly reduced computational runtimes compared to traditional neural network models. 
We are further able to scale the considered system sizes beyond most recent numerical studies, while keeping the accuracies high and computational costs reasonable~\cite{Czischek2022, Kalinowski2022, Samajdar2020, Merali2021}.
These observations promise the ability to study the scaling behavior of Rydberg atom arrays to large system sizes, allowing an in-depth exploration of the underlying phase diagram.
While such studies go beyond the scope of this proof-of-principle work, we leave it open for future follow-up works.

Our results show that the LPTF model performs similarly well in different phases of matter in the Rydberg system and accurately captures phase transitions.
While we focus on Rydberg atom arrays, the introduced approach can be applied to general quantum many-body systems, where complex-valued wave functions can be represented by adding a second output to the autoregressive network architecture as in~\cite{Hibat-Allah2020}.
While we expect the inclusion of complex phases to make the training process harder~\cite{Hibat-Allah2020}, modifications of the LPTF setup can be explored in future works to study more complex or larger qubit systems.
Such modifications include larger network models with more transformer cells, or higher embedding dimensions which increase the network expressivity~\cite{Zhai2021}.
Additionally, larger input patch sizes can be achieved by including multiple patched RNN and patched TF components in the LPTF architecture, which successively reduce the sub-patch sizes.
We further expect that the performance of LPTFs can be enhanced with a data-based initialization, as discussed in~\cite{Czischek2022, Bennewitz2022}.

Our results and possible future improvements promise high-quality representations of quantum states in various models and phases of matter at affordable computational costs.
This prospect proposes significant advances in the modeling of quantum many-body systems, promising insightful follow-up works exploring new physical phenomena.
%
%
\section{Methods}
\subsection{Rydberg atom arrays}
We apply our numerical methods on Rydberg atom arrays as an example for qubit systems.
In state-of-the-art experiments, Rydberg atoms are individually addressed via optical tweezers that allow for precise arrangements on arbitrary lattices in up to three dimensions~\cite{Endres2016, Ebadi2021, Scholl2021, Barredo2018}.
Fluorescent imaging techniques are then used to perform projective measurements in the Rydberg excitation basis.
Such accurate and well-controlled experimental realizations are accompanied by intensive numerical investigations, which have unveiled a great variety of phases of matter, separated by quantum phase transitions, in which Rydberg atom systems can be prepared~\cite{Samajdar2021, Ebadi2021, Samajdar2020, Kalinowski2022, Miles2021}.
The atoms on the lattice interact strongly via the Rydberg many-body Hamiltonian in \Eq{Hamiltonian}~\cite{Jaksch2000, Lukin2001}.
The Rydberg blockade radius $R_{\mathrm{b}}$ defines a region within which simultaneous excitations of two atoms are penalized.

The ground states of this Rydberg Hamiltonian are fully described by positive, real-valued wavefunctions so that the outcomes of measurements in the Rydberg occupation basis provide complete information about ground state wavefunctions~\cite{Endres2016, Ebadi2021, Xu2021}.
We can thus model ground state wavefunctions with real-valued neural network model architectures~\cite{Hibat-Allah2020, Carrasquilla2021}.
In this work, we choose $\Omega=1$ and describe the system in terms of the detuning $\delta$ and the Rydberg blockade radius $R_{\mathrm{b}}$.
We further consider square lattices of $N=L\times L$ atoms with lattice spacing $a=1$ and open boundary conditions.
%
\subsection{Recurrent neural network quantum states}
Recurrent neural networks (RNNs) are generative network architectures that are optimized to deal with sequential data~\cite{Hochreiter1997, Cho2014}.
They naturally encode a probability distribution and enable efficient sample data generation.
As illustrated in \Fig{1}\textbf{b},\textbf{d}, the RNN input is given by individual elements $\sigma_i$ of dimension $d_{\mathrm{I}}$ from a given data sequence $\boldsymbol{\sigma}$, and a hidden state $\boldsymbol{h}_i$ of dimension $d_{\mathrm{H}}$.
We use the initial input states $\boldsymbol{\sigma}_0=\boldsymbol{0}$ and $\boldsymbol{h}_0=\boldsymbol{0}$.
Throughout this work, we choose $d_{\mathrm{H}}=128$.
The input is processed in the RNN cell, where non-linear transformations defined via variational parameters $\mathcal{W}$ are applied.
Here we use the Gated Recurrent Unit (GRU)~\cite{Cho2014} as RNN cell, which is applied at each iteration with shared weights~\cite{Hochreiter1997}.

We then apply two fully connected projection layers on the hidden state, the first followed by a rectified linear unit (ReLU) activation function and the second followed by a softmax activation function (not layers are not shown in \Fig{1}\textbf{d}). 
This setup generates an output vector of dimension $d_{\mathrm{O}}$ which is interpreted as a probability distribution over all possible output values~\cite{Hibat-Allah2020}.
The hidden state configuration is propagated over the input sequence encoding information of previous inputs and generating a memory effect.
This setup conditions the output probability on all previous sequence elements,
$p_{\mathrm{RNN}}\left(\sigma_{i+1}\left|\sigma_{i},\dots,\sigma_1;\mathcal{W}\right.\right)$, from which an output state is sampled.
Here, we use this output as the input element $\sigma_{i+1}$ of the next iteration, running the network in an autoregressive manner.
In this case, the joint probability of the generated sequence is given by $p_{\mathrm{RNN}}\left(\boldsymbol{\sigma};\mathcal{W}\right)=\prod_ip_{\mathrm{RNN}}\left(\sigma_i\left|\sigma_{i-1},\dots,\sigma_1;\mathcal{W}\right.\right)$~\cite{Hibat-Allah2020}.

To use the RNN as a wavefunction ansatz to represent quantum states, we consider the quantum system as a sequence of qubits, sampling the state of one qubit at a time and using it as the input in the next RNN iteration~\cite{Hibat-Allah2020, Carrasquilla2021, Dawid2022}.
The hidden state propagation captures correlations in the qubit system by carrying information about previously sampled qubit configurations.
We then interpret the probability distribution encoded in the RNN as the squared wavefunction amplitude of the represented quantum state, $p_{\mathrm{RNN}}\left(\boldsymbol{\sigma};\mathcal{W}\right)=\left|\langle\boldsymbol{\sigma}\left|\Psi_{\mathcal{W}}\right.\rangle\right|^2=\left|\Psi_{\mathrm{RNN}}\left(\boldsymbol{\sigma};\mathcal{W}\right)\right|^2$.
This ansatz can model the complete information of ground states in the considered Rydberg Hamiltonian, \Eq{Hamiltonian}.
Samples generated from the RNN then correspond to outcomes of projective measurements in the computational basis and can be used to estimate expectation values of general observables $\hat{\mathcal{O}}$~\cite{Hibat-Allah2020, Carrasquilla2021, Dawid2022, Melko2019, Carleo2017, Torlai2018},
\begin{align}
	\nonumber
	\langle\Psi_{\mathcal{W}}\left|\mathcal{\hat{O}}\right|\Psi_{\mathcal{W}}\rangle &= \sum_{\left\{\boldsymbol{\sigma},\boldsymbol{\sigma}'\right\}}\Psi_{\mathrm{RNN}}^*\left(\boldsymbol{\sigma};\mathcal{W}\right)\Psi_{\mathrm{RNN}}\left(\boldsymbol{\sigma}';\mathcal{W}\right)\\
	\nonumber
	&\hphantom{=}\times\langle\boldsymbol{\sigma}\left|\hat{\mathcal{O}}\right|\boldsymbol{\sigma}'\rangle\\
	\nonumber
	&= \sum_{\left\{\boldsymbol{\sigma}\right\}}\left|\Psi_{\mathrm{RNN}}\left(\boldsymbol{\sigma};\mathcal{W}\right)\right|^2\mathcal{O}_{\mathrm{loc}}\left(\boldsymbol{\sigma};\mathcal{W}\right)\\
	&\approx \frac{1}{N_{\mathrm{s}}}\sum_{\substack{\boldsymbol{\sigma}_s\propto\\ p_{\mathrm{RNN}}\left(\boldsymbol{\sigma};\mathcal{W}\right)}}\mathcal{O}_{\mathrm{loc}}\left(\boldsymbol{\sigma}_s;\mathcal{W}\right),
	\label{eq:1a}
\end{align}
where we introduce the local observable, 
\begin{align}
	\nonumber
	\mathcal{O}_{\mathrm{loc}}\left(\boldsymbol{\sigma}_s;\mathcal{W}\right)&=\frac{\left\langle\boldsymbol{\sigma}_s\left|\hat{\mathcal{O}}\right|\Psi_{\mathcal{W}}\right\rangle}{\left\langle\boldsymbol{\sigma}_s\left|\Psi_{\mathcal{W}}\right.\right\rangle}\\	
	&= \sum_{\left\{\boldsymbol{\sigma}'\right\}}\left\langle \boldsymbol{\sigma}_s\left|\hat{\mathcal{O}}\right|\boldsymbol{\sigma}'\right\rangle\frac{\Psi_{\mathrm{RNN}}\left(\boldsymbol{\sigma}';\mathcal{W}\right)}{\Psi_{\mathrm{RNN}}\left(\boldsymbol{\sigma}_s;\mathcal{W}\right)}.
	\label{eq:1}
\end{align}
This local observable is evaluated and averaged over $N_{\mathrm{s}}$ samples $\boldsymbol{\sigma}_s$ generated from the RNN.
To find the ground state representation of a given Hamiltonian in the RNN, we use a gradient descent training algorithm to minimize the energy expectation value $\langle E\rangle = \langle\hat{\mathcal{H}}\rangle$, which can be similarly evaluated using samples from the RNN as stated in \Eq{E_exp} and \Eq{E_loc}~\cite{Hibat-Allah2020, Carrasquilla2021, Dawid2022, Melko2019, Becca2017}.
We train the RNN using the Adam optimizer with parameters $\beta_1=0.9$, $\beta_2=0.999$, and learning rate $\Delta=0.0005$.

The GRU cell has three internal weight matrices of dimension $d_{\mathrm{I}}\times d_{\mathrm{H}}$, with input dimension $d_{\mathrm{I}}$ and hidden unit dimension $d_{\mathrm{H}}$, and three internal weight matrices of dimension $d_{\mathrm{H}}\times d_{\mathrm{H}}$.
It furthermore has six internal bias vectors of size $d_{\mathrm{H}}$, and we add two fully connected layers with weight matrices of dimension $d_{\mathrm{H}}\times d_{\mathrm{H}}$ and $d_{\mathrm{H}}\times d_{\mathrm{O}}$ and biases of size $d_{\mathrm{H}}$ and $d_{\mathrm{O}}$, respectively, to obtain the desired RNN output vector with output dimension $d_{\mathrm{O}}$~\cite{Cho2014, Hibat-Allah2020}.
Single-qubit inputs give $d_{\mathrm{I}}=1$ and $d_{\mathrm{O}}=2$ as we use a one-hot encoded output.
Together with $d_{\mathrm{H}}=128$ as chosen in this work, this leads to a total of 
\begin{align}
	\nonumber
	&\hphantom{=}3\left(d_{\mathrm{I}}\times d_{\mathrm{H}}\right)+ 4\left(d_{\mathrm{H}}\times d_{\mathrm{H}}\right) + 7d_{\mathrm{H}} + \left(d_{\mathrm{H}}\times d_{\mathrm{O}}\right) + d_{\mathrm{O}} \\
	&= 67{,}074
	\label{eq:RNN}
\end{align}
trainable network parameters.
\subsection{Transformer quantum states}
Transformer (TF) models can be applied to sequential data similarly to RNNs.
Such models do not include a recurrent behavior but are based on self-attention, which provides access to all elements in the sequence and enables the dynamical highlighting of salient information.
We use the TF model as introduced in~\cite{Vaswani2017} and restrict it to the encoder part only~\cite{Cha2022, Sharir2022, Viteritti2022, Zhang2023}.

As illustrated in \Fig{1}\textbf{e}, the TF model first embeds the given input vector.
This embedding corresponds to a linear projection of the input vector of dimension $d_{\mathrm{I}}$ to a vector of embedding dimension $d_{\mathrm{H}}$ with trainable parameters $W^{\mathrm{I}}$.
As a next step, the positional encoding matrix is evaluated and added to the embedded input vector to include information about the positions of the input elements in the sequence~\cite{Vaswani2017}.
To keep this information when propagating the signal through the TF structure, the overall embedding dimension $d_{\mathrm{H}}$ of internal states is conserved.
Throughout this work, we choose $d_{\mathrm{H}}=128$.

The embedded input with positional encoding is passed to the transformer cell, where the query, key, and value matrices are generated, and multi-headed self-attention is applied~\cite{Vaswani2017}.
We use a masked self-attention mechanism to ensure the TF model is autoregressive, like the RNN.
The output vector of the masked self-attention is then added to the embedded input state, and the sum is normed before being fed into two feed-forward layers.
The normalization is a trained process to improve the network training stability and adds $2d_{\mathrm{H}}$ trainable parameters.
Similar to~\cite{Vaswani2017}, we apply one feed-forward layer with a ReLU activation function, followed by a linear feed-forward layer, where the weights and biases of both layers are trainable.
The first feed-forward layer projects the input into a vector of size $d_{\mathrm{FF}}=2048$, while the second feed-forward layer projects it back to size $d_{\mathrm{H}}$.
The output of the feed-forward layers is again added to the output vector of the self-attention cell, and the sum is normalized, see \Fig{1}\textbf{e}.

The entire transformer cell, including the self-attention and feed-forward layers, as well as the add-and-norm operations, can be applied multiple times independently to improve the network expressivity.
We obtain satisfying results with $T=2$.

To represent quantum states similarly to the RNN ansatz, we apply two fully connected layers with trainable weights to the output of the transformer cell.
The first layer conserves the dimension $d_{\mathrm{H}}$ and is followed by a ReLU activation function. 
The second layer projects the output to a vector of output dimension $d_{\mathrm{O}}$ and is followed by a softmax activation function.
These two layers are not shown in the diagram in \Fig{1}\textbf{e}.
After the softmax activation function, the output can be treated the same way as the RNN output, and it can be interpreted as a probability distribution from which the next qubit state in the sequence can be sampled~\cite{Vaswani2017}.
We train the TF model the same way as the RNN, using the Adam optimizer to minimize the energy expectation value via gradient descent.
The energy expectation value is obtained in the same way as in the RNN, see \Eq{E_exp} and \Eq{E_loc}, and we choose the same values as in the RNN approach for all hyperparameters involved in the training process.
\subsubsection{Positional encoding}
Since the TF model does not include a recurrent behavior as the RNN, it does not provide any information about the order of the sequence elements by default.
A positional encoding algorithm is used to include information about the position of each input.
We use the algorithm as proposed in~\cite{Vaswani2017}, which creates a matrix of dimension $L\times d_{\mathrm{H}}$ with sequence length $L$ and embedding dimension $d_{\mathrm{H}}$.
The individual elements are calculated via
\begin{align}
	P\left(l, 2i\right) &= \mathrm{sin}\left[\frac{l}{10000^{2i/d_{\mathrm{H}}}}\right],\\
	P\left(l, 2i+1\right) &= \mathrm{cos}\left[\frac{l}{10000^{2i/d_{\mathrm{H}}}}\right],
\end{align}
with $0\leq l< L$ indexing the sequence elements, and $0\leq i<d_{\mathrm{H}}/2$ the column indices of the output matrix.
The resulting matrix is added to the embedded input element of the TF setup, which linearly projects the input vector to a vector of dimension $d_{\mathrm{H}}$ using trainable weights.
This operation gives each element a unique information about its position in the sequence.
\subsubsection{The self-attention mechanism}
\begin{figure}
	\includegraphics{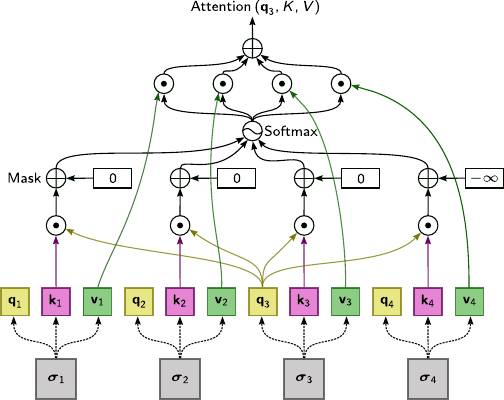}
	\caption{\textbf{Illustration of the attention mechanism applied to sequence element $\boldsymbol{\sigma}_3$.}
		Each sequence element $\boldsymbol{\sigma}_j$ is projected onto a query ($\boldsymbol{q}_j$), key ($\boldsymbol{k}_j$), and value ($\boldsymbol{v}_j$) vector, and the dot product of $\boldsymbol{q}_3$ is taken with each key vector $\boldsymbol{k}_j$, $j\in\left\{1,2,3,4\right\}$.
		Adding a mask matrix eliminates the influence of all elements later in the sequence by adding $m_{i,j>i}=-\infty$, while contributions of previously sampled sequence elements remain unchanged with $m_{i,j\leq i}=0$, ensuring the autoregressive behavior.
		A softmax activation function is applied to all signals before a dot product is taken with the corresponding value vectors.
		The sum of all resulting signals provides the attention outcome for the input $\boldsymbol{\sigma}_3$.
		This algorithm is the same as introduced in~\cite{Vaswani2017}.
	}
	\label{fig:5}
\end{figure}
The self-attention mechanism, as introduced in~\cite{Vaswani2017} and illustrated in \Fig{5}, projects each embedded sequence element $\boldsymbol{\sigma_i}$ of dimension $d_{\mathrm{H}}$ to a query vector $\boldsymbol{q}_i$, a key vector $\boldsymbol{k}_i$, and a value vector $\boldsymbol{v}_i$ of dimensions $d_{\mathrm{H}}$,
\begin{align}
	\boldsymbol{q}_i &= \sum_{l=1}^{d_{\mathrm{H}}}W_{i,l}^{\mathrm{q}}\sigma_{i,l},\ \ \ \boldsymbol{k}_i=\sum_{l=1}^{d_{\mathrm{H}}}W_{i,l}^{\mathrm{k}}\boldsymbol{\sigma}_{i,l},\ \ \ \boldsymbol{v}_i = \sum_{l=1}^{d_{\mathrm{H}}}W_{i,l}^{\mathrm{v}}\boldsymbol{\sigma}_{i,l},
\end{align}
with trainable weight matrices $W^{\mathrm{q}}$, $W^{\mathrm{k}}$, $W^{\mathrm{v}}$ of dimension $d_{\mathrm{H}}\times d_{\mathrm{H}}$.
Query, key, and value vectors of all input elements can be summarized in the corresponding matrices,
\begin{align}
	Q&=\left[\begin{matrix}\boldsymbol{q}_1\\ \vdots \\ \boldsymbol{q}_L\end{matrix}\right],\ \ \ K=\left[\begin{matrix}\boldsymbol{k}_1\\ \vdots \\ \boldsymbol{k}_L\end{matrix}\right],\ \ \ V=\left[\begin{matrix}\boldsymbol{v}_1\\ \vdots \\ \boldsymbol{v}_L\end{matrix}\right],
\end{align}
with sequence length $L$.

The attention mechanism then maps the queries and key-value pairs to an output for each sequence element, allowing for highlighting of connections to sequence elements with important information.
For each sequence element $\boldsymbol{\sigma}_i$, the dot product of the query vector $\boldsymbol{q}_i$ with the key vector $\boldsymbol{k}_j$ for all $j\in\left\{1,\dots,L\right\}$ is evaluated.
We then add a masking term $m_{i,j}$ to the signal, which is given by,
\begin{align}
	m_{i,j} &= \begin{cases}0\ \mathrm{if}\ i\leq j,\\ -\infty\ \mathrm{otherwise.}\end{cases}
	\label{eq:app1}
\end{align}
This ensures that the self-attention only considers previous elements in the sequence and does not look at later elements that still need to be determined in the autoregressive behavior.
Applying a softmax activation function to all signals after adding the mask ensures that the contributions of all later sequence elements with $m_{i,j}=-\infty$ are driven to zero.
We then take the dot product of each signal with the corresponding value vector $\boldsymbol{v}_j$ and sum all signals to generate the output of the attention mechanism.
The complete attention formalism can thus be summarized as,
\begin{align}
	\mathrm{Attention}\left(Q,K,V\right) &= \mathrm{softmax}\left(\frac{QK^{\mathrm{T}}}{\sqrt{d_{\mathrm{H}}/h}}+M\right)V,
	\label{eq:att}
\end{align}
where the mask matrix $M$ has entries $m_{i,j}$ as in \Eq{app1}.
Here we further use multi-headed attention, as discussed in~\cite{Vaswani2017}.
This approach linearly projects each query, key, and value vector to $h$ vectors with individually trainable projection matrices.
We thus end up with $h$ heads with modified query, key, and value vectors on which the attention mechanism is applied, where we choose $h=8$ throughout this work, as we find it to yield satisfying results.
The linear projection further reduces the dimension of the query, key, and value vectors to $d_{\mathrm{H}}/h$, so that the outputs of the individual heads can be concatenated to yield the total output dimension $d_{\mathrm{H}}$ of the attention mechanism.
We then scale the outcome of each query-key dot product with the factor $1/\sqrt{d_{\mathrm{H}}/h}$ in \Eq{att}~\cite{Vaswani2017}.
The output of the multi-headed attention formalism is given by,
\begin{align}
	\mathrm{Multihead}\left(Q,K,V\right) &= \mathrm{Concat}\left(\hat{y}_1,\dots,\hat{y}_h\right)W^{\mathrm{O}},\\
	\hat{y}_l\left(Q,K,V\right) &= \mathrm{Attention}\left(QW_l^{\mathrm{Q}}, KW_{l}^{\mathrm{K}}, VW_{l}^{\mathrm{V}}\right),
\end{align}
with output weight matrix $W^{\mathrm{O}}$ and query, key, and value weight matrices $W_l^{\mathrm{Q}}$, $W_l^{\mathrm{K}}$, and $W_{l}^{\mathrm{V}}$ for head $l$.

A TF model given an input of dimension $d_{\mathrm{I}}$ has an embedding matrix $W_{\mathrm{I}}$ of dimension $d_{\mathrm{I}}\times d_{\mathrm{H}}$, with embedding dimension $d_{\mathrm{H}}$.
The weight matrices in the multi-headed self-attention mechanism then have dimensions $d_{\mathrm{H}}\times d_{\mathrm{H}}/h$ for $W_l^{\mathrm{Q}}$, $W_{l}^{\mathrm{K}}$, and $W_l^{\mathrm{V}}$, and $d_{\mathrm{H}}\times d_{\mathrm{H}}$ for $W^{\mathrm{O}}$.
Each weight matrix also comes with a bias whose size equals the column-dimension.
The two feed-forward layers in the transformer cell have weight matrices of $d_{\mathrm{H}}\times d_{\mathrm{FF}}$ and $d_{\mathrm{FF}}\times d_{\mathrm{H}}$ with corresponding biases, and the two norm operations add $4d_{\mathrm{H}}$ trainable parameters.
The transformer cell, containing the attention mechanism, the feed-forward layers, and the norm operations, is applied $T$ times with independent variational parameters.
After the transformer cell we add two fully connected layers with weight matrices of dimensions $d_{\mathrm{H}}\times d_{\mathrm{H}}$ and $d_{\mathrm{H}}\times d_{\mathrm{O}}$ for output dimension $d_{\mathrm{O}}$.
Both layers come with corresponding biases.

Single-qubit inputs give $d_{\mathrm{I}}=1$ and $d_{\mathrm{O}}=2$, using one-hot encoded output. With $d_{\mathrm{H}}=128$, $d_{\mathrm{FF}}=2048$, and $T=2$, as chosen throughout this work, the TF architecture has a total of
\begin{align}
	\nonumber
	&\left(d_{\mathrm{I}}\times d_{\mathrm{H}}\right) +d_{\mathrm{H}} + T\left[4\left(d_{\mathrm{H}}\times d_{\mathrm{H}}\right)+9d_{\mathrm{H}}+2\left(d_{\mathrm{FF}}\times d_{\mathrm{H}}\right)\right.\\
	\nonumber
	&+\left.d_{\mathrm{FF}}\right]+\left(d_{\mathrm{H}}\times d_{\mathrm{H}}\right)+d_{\mathrm{H}}+\left(d_{\mathrm{H}}\times d_{\mathrm{O}}\right)+d_{\mathrm{O}}\\
	&= 1{,}203{,}074	
\label{eq:TF}
\end{align}
trainable variational parameters.
\subsection{Patched network models}
The bottleneck of the RNN and TF wavefunction ansatz is the iteration of the network cell over the entire qubit sequence.
This computationally expensive step needs to be done for each sample that is generated, as well as each time a wavefunction $\Psi\left(\boldsymbol{\sigma};\mathcal{W}\right)$ is calculated, which is required to evaluate non-diagonal observables, see \Eq{1}.
We reduce the number of iterations per network call by shortening the input sequence and in return increasing the dimension of the input vector.

As illustrated in \Fig{1}, for two-dimensional Rydberg atom arrays, we consider patches of $p$ qubits arranged in squares or rectangles.
We flatten these patches into binary input vectors of dimension $d_{\mathrm{I}}=p$.
This modification increases the network input dimension, which is, however, projected to the unaffected hidden state dimension in the RNN cell or the embedding dimension in the TF model.
Thus, the computational cost of evaluating the network cell is barely affected by the increased input dimension, but the shorter sequence length leads to significantly reduced computational runtimes.
In addition to this expected speed-up, we expect the patched network models to capture local correlations in the system with higher accuracy.
Neighboring qubit states are now considered at the same iteration, and their information is not encoded in the network state.

The network output uses one-hot encoding of the patched quantum states, so that the output vector is of dimension $d_{\mathrm{O}}=2^{p}$.
Each entry represents one possible state of the qubit patch, see \Fig{1}\textbf{d},\textbf{e}.
This output dimension, and with this the computational cost of evaluating the softmax function, thus scales exponentially with the patch size.
In this work, we only consider patches up to $p=2\times 2$ for the patched network models and introduce large, patched TFs to deal with larger patch sizes.

The patched RNN with $p=2\times 2$ has input dimension $d_{\mathrm{I}}=4$ and output dimension $d_{\mathrm{O}}=16$, so \Eq{RNN} leads to $70{,}032$ trainable network parameters.
For the same input and output dimension, the patched TF model has $1{,}203{,}406$ trainable network parameters, according to \Eq{TF}.
\subsection{Large, patched transformer models}
In the large, patched transformer (LPTF) model, we apply the TF network to a patch of $p$ qubits.
However, we abort the TF model in \Fig{1}\textbf{e} right after the transformer cell and do not apply the fully connected layers and the softmax activation function.
Instead, we use the generated output state of the transformer cell as an input hidden state to a patched RNN with the hidden-unit dimension matching the embedding dimension $d_{\mathrm{H}}$ of the TF model. 
This patched RNN breaks up the large input patch into smaller sub-patches of size $p_{\mathrm{s}}$, where we always choose a sub-patch size of $p_{\mathrm{s}}=2\times 2$ in this work.
We then use the patched RNN model to iteratively sample the quantum states of these sub-patches in the same way as when applying the patched RNN to the full system size.
The only difference is that the initial hidden state is provided by the TF output, $\boldsymbol{h}_{\mathrm{0}}=\boldsymbol{h}_{\mathrm{TF}}$, see \Fig{1}\textbf{f} for an illustration.

The total number of trainable network parameters in the LPTF setup is then given by a combination of \Eq{RNN} and \Eq{TF}, where the two fully connected layers at the TF output are subtracted,
\begin{align}
	\nonumber 
	&\underbrace{\left(p\times d_{\mathrm{H}}\right) + T\left[4\left(d_{\mathrm{H}}\times d_{\mathrm{H}}\right) + 9d_{\mathrm{H}} + 2\left(d_{\mathrm{FF}}\times d_{\mathrm{H}}\right) + d_{\mathrm{FF}}\right]}_{\mathrm{patched}\ \mathrm{TF}}\\
	+&\underbrace{3\left(p_{\mathrm{s}}\times d_{\mathrm{H}}\right) + 4\left(d_{\mathrm{H}}\times d_{\mathrm{H}}\right) + 7d_{\mathrm{H}} + \left(d_{\mathrm{H}}\times d_{\mathrm{O}}\right) + d_{\mathrm{O}}}_{\mathrm{patched}\ \mathrm{RNN}}.
\end{align}
We use the input dimension $d_{\mathrm{I}}=p$ for the patched TF and $d_{\mathrm{I}}=p_{\mathrm{s}}$ for the patched RNN, as well as the output dimension $d_{\mathrm{O}}=2^{p_{\mathrm{s}}}$.
In this work, we choose $p_{\mathrm{s}}=2\times 2$, which yields $1{,}256{,}208+128p$ trainable parameters with $d_{\mathrm{H}}=128$, $d_{\mathrm{FF}}=2048$, and $T=2$.
For the choice $p=8\times 8$, we thus get $1{,}264{,}400$ variational network parameters.
\subsection{Computational complexity}
The process of finding ground state representations with ANNs can be divided into two steps, the generation of samples from the network and the evaluation of energy expectation values according to \Eq{1a} and \Eq{1} in each training iteration.
We start with analyzing the sample complexity for the different network architectures.
As we choose the hidden and the embedding dimension $d_{\mathrm{H}}$ fixed and equal for all architectures, we consider it as a constant in the complexity analysis.

Generating a single sample $\boldsymbol{\sigma}$ from $p_{\mathrm{RNN}}\left(\boldsymbol{\sigma};\mathcal{W}\right)$ encoded in an RNN requires $N$ executions of the RNN cell, leading to a computational cost of $\mathcal{O}\left(N\right)$.
By considering the patched RNN, we reduce the sequence length from $N$ to $N/p$, so that only $N/p$ RNN cells are evaluated.
However, the output dimension in this case is $2^p$, so each evaluation of the RNN cell requires $2^p$ products to evaluate the outcome probability distribution.
This leads to an overall sampling complexity of $\mathcal{O}\left(N/p2^p\right)$ for the patched RNN.

In order to generate a single sample $\boldsymbol{\sigma}$ from $p_{\mathrm{TF}}\left(\boldsymbol{\sigma};\mathcal{W}\right)$ encoded in the TF network, we similarly need to evaluate the transformer cell $N$ times.
However, the attention algorithm itself requires the computation of $N$ multiplications that need to be evaluated in each pass through the network.
Additionally, the transformer cell contains a projection of the embedded state to a vector of size $d_{\mathrm{FF}}\gg d_{\mathrm{H}}$, which requires significantly more multiplication operations than an RNN cell evaluation.
Thus, drawing a sample from the TF model comes at computational complexity $\mathcal{O}\left(N\left[N+d_{\mathrm{eff}}\right]\right)$.
When introducing the patched TF model, we similarly reduce the sequence length to $N/p$, so that the full network needs to be evaluated $N/p$ times and the attention mechanism requires $N/p$ multiplications.
Also here the output scales as $2^p$, leading to a computational cost of $\mathcal{O}\left(N/p\left[N/p+d_{\mathrm{FF}}+2^p\right]\right)$.

In the LPTF, the transformer cell is followed by a patched RNN with $p/p_{\mathrm{s}}$ cells.
Since each LPTF iteration requires the evaluation of one such RNN, we evaluate the transformer cell and $p/p_{\mathrm{s}}$ RNN cells $N/p$ times to generate a single sample $\boldsymbol{\sigma}$.
While the TF network output is kept at embedding dimension $d_{\mathrm{H}}$, the RNN output is of dimension $2^{p_{\mathrm{s}}}$, leading to a computational complexity of 
\begin{align}
\nonumber
\mathcal{O}&\left(\frac{N}{p}\left[\frac{N}{p}+d_{\mathrm{FF}}\right]+\frac{N}{p}\frac{p}{p_{\mathrm{s}}}2^{p_{\mathrm{s}}}\right)\\
&=\mathcal{O}\left(\frac{N^2}{p^2}+\frac{N}{p}d_{\mathrm{FF}}+\frac{N}{p_{\mathrm{s}}}2^{p_{\mathrm{s}}}\right).
\end{align}
This shows a significant reduction of the sampling complexity compared to the patched TF model and explains the observed efficiency of our introduced LPTF architecture. 

Next, we consider the complexity of evaluating energy expectation values.
While the evaluation of the diagonal part is given by a linear average over all $N_{\mathrm{s}}$ samples and thus scales linearly with the system size $N$, evaluating the off-diagonal part for each sample $\boldsymbol{\sigma}_s$ according to \Eq{1} requires the evaluation of $\Psi\left(\boldsymbol{\sigma}';\mathcal{W}\right)$ for all $\boldsymbol{\sigma}'$ corresponding to $\boldsymbol{\sigma}_s$ with one atom flipped.
This leads to $N$ evaluations of $\Psi\left(\boldsymbol{\sigma}';\mathcal{W}\right)$ for each sample, which is obtained by passing $\boldsymbol{\sigma}'$ through the network architecture and obtaining the output probability $p_{\mathrm{RNN}}\left(\boldsymbol{\sigma}';\mathcal{W}\right)$ or $p_{\mathrm{TF}}\left(\boldsymbol{\sigma}';\mathcal{W}\right)$.

Thus, the patched RNN with $N/p$ network cells needs to be evaluated $N$ times for each sample, leading to a computational complexity of $\mathcal{O}\left(N_{\mathrm{s}}2^p N^2/p\right)$ for obtaining the energy expectation value.
Similarly, the patched TF model with $N/p$ iterations and $N/p$ multiplications in the attention mechanism is evaluated $N$ times for each sample, leading to a computational cost of $\mathcal{O}\left(N_{\mathrm{s}}\left[N^3/p^2+d_{\mathrm{FF}}N^2/p+2^p N^2/p \right]\right)$.
For the LPTF we accordingly obtain $\mathcal{O}\left(N_{\mathrm{s}}\left[N^3/p^2+d_{\mathrm{FF}}N^2/p+2^{p_{\mathrm{s}}}N^2/p_{\mathrm{s}}\right]\right)$.
The required memory scaling behaves similarly for the discussed network architectures.
This scaling can be reduced using optimized implementation algorithms as discussed in the next section.
While the TF and LPTF show a worse scaling than the RNN, the evaluation of the off-diagonal energy terms can be parallelized for these two models.
Since no autoregressive sampling is required for this task, all $N/p$ masked self-attention layers can be evaluated in parallel, significantly reducing the computational runtime.
This parallelization is not possible for the RNN due to its recurrent nature, where the hidden state needs to be evaluated for each individual RNN cell before it can be passed to the next iteration.

The generation of samples with QMC is of computational complexity $\mathcal{O}\left(VN\right)$, with average interaction strength $V$ over the system.
The energy estimation also scales as $\mathcal{O}\left(VN\right)$ and only needs to be done at the end of the run after all samples have been generated, which is in contrast to the neural network approach that requires the evaluated energy in each training iteration~\cite{Merali2021}.
While QMC thus shows much more promising computational complexity than all three neural network methods when considering the scaling to large system sizes, we observe that it requires longer computational runtimes than the LPTF for the system sizes considered in this work.
Considering the uncertainties of expectation values, we find that QMC simulations require far more samples ($N_{\mathrm{s}}=7\times 10^4$ in \Fig{2}, \Fig{3}, and $N_{\mathrm{s}}=7\times 10^5$ in \Fig{4}) than the ANN approaches ($N_{\mathrm{s}}=512$).
However, the ANN approaches require the generation of $N_{\mathrm{s}}$ samples in each training iteration, so that overall more samples are generated.

The higher uncertainties in the QMC simulations are caused by correlations in the generated sample chains, where autocorrelation times grow with increasing system sizes.
Furthermore, ergodicity in the QMC sampling process is not guaranteed for large systems~\cite{Merali2021}.
These problems do not arise in the exact and independent autoregressive sampling of the neural network algorithms, explaining the lower uncertainties observed for smaller sample sizes.
At the same time, these observations limit accurate QMC simulations to small system sizes.
\subsection{Implementation details}
We train the network to find the ground state of the Rydberg Hamiltonian by minimizing the energy expectation value, which we evaluate using \Eq{1a} and \Eq{1} with the Hamiltonian operator $\hat{\mathcal{H}}$,
\begin{align}
	\nonumber
	\langle E\rangle &= \left\langle\Psi_{\mathcal{W}}\left|\hat{\mathcal{H}}\right|\Psi_{\mathcal{W}}\right\rangle\\
	&= \sum_{\left\{\boldsymbol{\sigma},\boldsymbol{\sigma}'\right\}}\Psi^*\left(\boldsymbol{\sigma};\mathcal{W}\right)\Psi\left(\boldsymbol{\sigma}';\mathcal{W}\right)\left\langle\boldsymbol{\sigma}\left|\hat{\mathcal{H}}\right|\boldsymbol{\sigma}'\right\rangle,
\end{align}
where $\Psi\left(\boldsymbol{\sigma};\mathcal{W}\right)$ denotes a wavefunction encoded in one of the discussed network models.
To optimize the variational network parameters, we use the gradient descent algorithm the same way as discussed in~\cite{Hibat-Allah2020, Carrasquilla2021}, with gradients
\begin{align}
	\partial_{\mathcal{W}_i}E \approx \frac{2}{N_{\mathrm{s}}}\sum_{s=1}^{N_\mathrm{s}}\partial_{\mathcal{W}_i}\Psi^*\left(\boldsymbol{\sigma}_{s};\mathcal{W}\right)\left[E_{\mathrm{loc}}\left(\boldsymbol{\sigma}_{s};\mathcal{W}\right)-\langle E\rangle\right],
\end{align}
and local energy 
\begin{align}
E_{\mathrm{loc}}\left(\boldsymbol{\sigma}_s;\mathcal{W}\right) &= \sum_{\left\{\boldsymbol{\sigma}'\right\}}\left\langle\boldsymbol{\sigma}_s\left|\hat{\mathcal{H}}\right|\boldsymbol{\sigma}'\right\rangle\frac{\Psi\left(\boldsymbol{\sigma}';\mathcal{W}\right)}{\Psi\left(\boldsymbol{\sigma}_s;\mathcal{W}\right)}.
\end{align}
The training process requires the evaluation of the gradients of $\Psi\left(\boldsymbol{\sigma};\mathcal{W}\right)$.
To reduce the necessary amount of memory, we always generate a batch of $N_{\mathrm{s}}$ samples from the network without evaluating the gradients.
We then pass each sample through the network again to obtain the wavefunction amplitude with the corresponding gradients.
This approach requires $2N_{\mathrm{s}}$ network passes instead of $N_{\mathrm{s}}$, but evaluating gradients on a given input sequence is less memory-consuming than evaluating gradients on an autoregressive process in PyTorch~\cite{Paszke2019}.
We can further reduce the required memory by dividing the total batch of $N_{\mathrm{s}}$ samples into mini batches of $K$ samples each, which are evaluated in separate processes.
This reduces the memory scaling by a factor $K/N_{\mathrm{s}}$ per pass, while requiring $N_{\mathrm{s}}/K$ network passes instead of one.
Thus, the smaller we choose the mini batches, the less memory is required, but the longer the computational runtime.
If not stated otherwise, we choose $K=256$ in this work.

Considering the off-diagonal term in the Rydberg Hamiltonian, its contribution to the energy expectation value is given by $E_{\mathrm{off}}=\sum_{i=1}^N\left\langle\hat{\sigma}_i^x\right\rangle$.
Thus, calculating the local energy $E_{\mathrm{loc}}\left(\boldsymbol{\sigma}_s;\mathcal{W}\right)$ using \Eq{1} requires for each sampled state $\boldsymbol{\sigma}_s$ to evaluate $\Psi\left(\boldsymbol{\sigma}';\mathcal{W}\right)$ for all $\boldsymbol{\sigma}'$ that correspond to the sampled state with one atom flipped from the ground to the excited state or vice versa~\cite{Hibat-Allah2020, Carrasquilla2021}.
Instead of passing $N$ states through the network for each sample we generate, we can reduce the required memory and accelerate our algorithm by splitting the atom sequence into $D$ equally sized parts.
In each part, the wavefunction amplitude is evaluated for the states where each of the $N/D$ atoms is flipped.
When calculating the wavefunction amplitudes using sequential networks, the flipping of one atom only affects the calculation for atoms that appear later in the sequence.
We store the network outcome for the original state $\boldsymbol{\sigma}_s$ after each $N/D$ atoms and evaluate each group starting from this stored value.
We then only need to pass the sequence from the first atom of the group to the last atom in the system to the network.
This ansatz can be used since no gradient needs to be evaluated on this off-diagonal term and corresponds to using mini batches of atoms.
This way we reduce the amount of required memory by a factor $D/N$ per pass, while requiring $N/D$ network passes instead of only one.
While we now evaluate the network $N/D$ times, the sequence length at iteration $d$ is reduced from $N/p$ to $\left(N-\left[d-1\right]D\right)/p$, leading to an accelerated evaluation of the local energies since the input sequence length is reduced in most cases.
In this work, we always split the sequence of atoms into $D=N/p$ parts, with $p$ the patch size in the patched network models.

We base our simulations on PyTorch~\cite{Paszke2019} and NumPy~\cite{Harris2020}, and use Matplotlib~\cite{Hunter2007} to visualize our results.
\section{Data availability}
All presented data can be reproduced with the publicly available source code.
It is further available upon request to the corresponding author.
\section{Code availability}
The source code used to generate the data in this work is available on \url{https://github.com/APRIQuOt/VMC_with_LPTFs.git}.
It is based on PyTorch~\cite{Paszke2019} and NumPy~\cite{Harris2020} and we used Matplotlib~\cite{Hunter2007} for visualizing our results.
%
\section{Acknowledgments}
We thank J. Carrasquilla, R.G. Melko, M. Reh, M.S. Moss, and E. Inack for fruitful discussions and feedback.
We are grateful for support on the quantum Monte Carlo simulations by E. Merali.
This research was enabled in part by support provided by the Digital Research Alliance of Canada (alliancecan.ca).
\section{Author contributions}
The fundamental ideas of the introduced approach were developed by K. Sprague who further implemented and organized the used Python code.
S. Czischek used the provided code to obtain the presented results with support by K. Sprague.
The manuscript was written by S. Czischek with valuable feedback by K. Sprague.
\section{Competing interests}
The authors declare no competing interests.
S. Czischek is a guest editor for \textit{Communications Physics}, but was not involved in the editorial review of, or the decision to publish this article.
\end{document}